\documentclass[pra,aps,reprint,superscriptaddress]{revtex4-1}

\usepackage{amsmath}    
\usepackage{graphicx}   
\usepackage{verbatim}   
\usepackage{color}      
\usepackage{subfigure}  
\usepackage{hyperref}   

\usepackage{braket}

\newcommand{\tra}[2]{\text{tr}_{#1}\left\lbrace#2\right\rbrace} 

\begin{document}

\title{Initial System-Environment Correlations via the Transfer Tensor Method}
\author{Maximilian Buser}
\affiliation{Institut f\"ur Theoretische Physik, Technische Universit\"at Berlin, Hardenbergstr. 36, D-10623 Berlin, Germany}
\affiliation{Department of Chemistry, Massachusetts Institute of Technology, 77 Massachusetts Avenue, Cambridge, Massachusetts 02139, USA}
\author{Javier Cerrillo}
\email{cerrillo@tu-berlin.de}
\affiliation{Institut f\"ur Theoretische Physik, Technische Universit\"at Berlin, Hardenbergstr. 36, D-10623 Berlin, Germany}
\author{Gernot Schaller}
\affiliation{Institut f\"ur Theoretische Physik, Technische Universit\"at Berlin, Hardenbergstr. 36, D-10623 Berlin, Germany}
\author{Jianshu Cao}
\affiliation{Department of Chemistry, Massachusetts Institute of Technology, 77 Massachusetts Avenue, Cambridge, Massachusetts 02139, USA}

\date{\today}

\begin{abstract}
Open quantum systems exhibiting initial system-environment correlations are notoriously difficult to simulate.
We point out that given a sufficiently long sample of the exact short-time evolution of the open system dynamics, 
one may employ transfer tensors for the further propagation of the reduced open system state. 
This approach is numerically advantageous and allows for the simulation of quantum correlation functions in 
hardly accessible regimes. 
We benchmark this approach against analytically exact solutions and exemplify it with the calculation of emission 
spectra of multichromophoric systems as well as for the reverse temperature estimation from simulated spectroscopic data. 
Finally, we employ our approach for the detection of spectral signatures of electromagnetically-induced transparency in open three-level systems.
\end{abstract}

\maketitle

%
%
%
%
\section{Introduction}

The simulation of reduced open quantum systems is usually facilitated by the assumption of at least an initial product state between the system and its environment at some initial point in time~\cite{breuer2002theory, weiss2012quantum}. 
However, numerous questions crucially require to account for initial correlations 
between the open system of interest and its environment.
Apart from witnessing or certifying their existence~\cite{laine2010a,rodriguez_rosario2012a,chaudhry2013role,ringbauer2015a,vinjanampathy2015a,chen_goan}, 
they find very concrete applications such as the correct simulation of emission spectra of molecular systems~\cite{cao2015forster_1,cao2015forster_2,cao2015forster_3} or their thermodynamic effect for the extractable work of thermal machines \cite{Alicki2013,Perarnau2015}. 
Although it has recently been pointed out that correlations may play a relevant role already in the weak system-environment coupling regime~\cite{chaudhry2013role}, 
it is in situations of strong coupling to a non-Markovian environment where initial correlation effects are expected to be most significant. 
In this regime, it becomes necessary to resort to specialized treatments, see e.g. Refs.~\cite{grabert1988,pechukas_1994,gaspard_2003,iles_smith2014a,chen_goan,halimeh_vega,strasberg2016a}, 
at the expense of enhanced computational effort. 

Here, methods such as the stochastic Liouville-von Neumann equations (SLE)~\cite{stockburger2002,shi2015} or the hierarchy of equations of motion 
technique (HEOM)~\cite{tanimura_1989,tanimura_2014} stand out. 
SLE is based on the sampling of environmental trajectories in terms of stochastic realizations. 
The number of required trajectories generally scales with the length of the simulation time rather than with the strength of the coupling to the environment,
its spectral density or its temperature, and it is therefore well suited for short but accurate simulations. 
In contrast, HEOM scales with the number of exponential functions required to approximate the environmental correlation functions 
and with the coupling strength. 
For this reason, it is best suited for high-temperature baths and Lorentzian spectral densities.

\begin{figure}[ht]
\includegraphics[width=\columnwidth,clip=true]{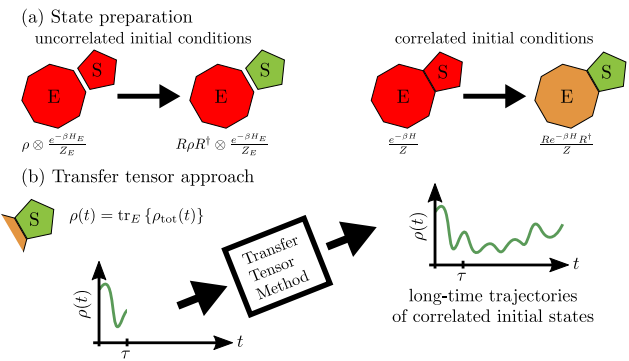}
\caption{\label{fig0}
Sketch of the considered setup (a) and the employed transfer tensor approach for the simulation of correlated initial states (b). 
(a) Correlated initial states are obtained by means of preparative transformations $R \bullet R^\dagger$ of the thermal steady 
state of the system-environment compound. 
Even though $R \bullet R^\dagger$ acts on the open system subspace only, the environment is indirectly affected by this 
transformation when correlations are present. 
(b) Short-time trajectories of the correlated reduced open system state are computed by means of HEOM or SLE. 
These short-time trajectories contain key information about the environmental influence and the effect of initial 
correlations onto the open system dynamics. 
This information is extracted by means of TTM and used for an efficient long-time propagation of the correlated initial state.
}
\end{figure}

Strategies to circumvent the unfavorable scaling have been put forward~\cite{stockburger16}. 
In this context, the transfer tensor method (TTM)~\cite{cerrillo_tt} provides a powerful tool to extend simulations 
to very long times~\cite{rosenbach_2016}. It can be interpreted as a discrete representation of the Nakajima-Zwanzig equation and has additionally found applications to mixed quantum-classical methods~\cite{kananenka2016accurate}.
While originally transfer tensors are obtained from simulations of separable initial states, here we demonstrate 
how they can be correctly applied to account for a large class of initial system-environment correlations.
This enables us to efficiently compute two-time correlation functions in the steady state such as absorption and emission spectra, 
as illustrated in Fig.~\ref{fig0}. 
The class of initial states that are accessible with the method can be regarded as preparative transformations of 
the global thermal system-environment state. 
We restrict ourselves to preparative transformations that affect only the system but are otherwise general, 
including e.g. a measurement performed on the open system or system-local unitary transformations. 

In the following, we first illustrate in Sec.~\ref{SEC:quantum_regression} the breakdown of the quantum regression theorem by exploring how within a HEOM simulation the effect 
of initial correlations becomes increasingly relevant in the strong-coupling and slow-environment regime. 
In Sec.~\ref{SEC:init_corr_ttm}, we demonstrate that correlated initial states can be treated with TTM, where a successful benchmark of the 
joint TTM-HEOM approach is also presented. 
In Sec.~\ref{SEC:emission_spectra}, we apply the proposed TTM approach for correlated initial states to the calculation of emission spectra 
of multichromophoric systems in previously hardly accessible regimes. 
Additionally, a detailed balance relation between absorption and emission spectra is employed for successful temperature estimation from simulated spectra. 
Finally, we use the TTM-SLE approach for the detection of spectral signatures of electromagnetically induced transparency 
in an open three-level system in Sec.~\ref{SEC:darkstates}.

%
%
%
%
\section{Breakdown of the quantum regression theorem}\label{SEC:quantum_regression}

We illustrate that the effect of initial correlations becomes especially relevant for systems located beyond the weak-coupling regime or systems 
that feature large correlation times of the environment. 
This is formally equivalent to the breakdown of the quantum regression theorem. 
We consider systems that are described in terms of a total Hamiltonian $H$, 
comprised of a system ($S$), interaction ($SE$) and environmental ($E$) part, $H=H_S+H_{SE}+H_E$. 
The correlated initial states of interest are obtained from the thermal system-environment state, 
characterized by an inverse temperature $\beta$, and subsequently transformed by a preparative procedure $R \bullet R^\dagger$, 
\begin{align}
\rho_\text{tot}(t_0)=R \frac{e^{-\beta H}}{Z} R^\dagger, \label{eq_init_state}
\end{align}
where $R$ only affects the system.
We will use a notation where $\rho_\text{tot}(t)$ denotes the total density matrix, and where the partition functions 
corresponding to the total system thermal state $Z$ and the thermal environmental state $Z_E$ are given by 
$Z_{(E)}=\text{tr}\left\lbrace e^{-\beta H_{(E)}} \right\rbrace$. 
The reduced state describing the open system of interest is then obtained from the total system state by tracing 
out the environmental degrees of freedom, $\rho(t)=\text{tr}_E\left\lbrace \rho_\text{tot}(t) \right\rbrace$. 
Regarding the preparative procedures, they may be of the form of a projective measurement performed on the open system, 
so that $R=R^\dagger=P_o/\sqrt{\text{tr}\left\lbrace P_o e^{-\beta H}/Z\right\rbrace}$ with $P_o=P_o^\dagger$ being 
the projector associated to the respective measurement outcome. 
Alternatively, $R \bullet R^\dagger$ might also denote a unitary transformation that acts on the open system subspace only. 
Any admissible transformation for a density matrix is suitable as long as it is applied locally on the system (Kraus map) \cite{Kraus1983}.
From now on, we will set the initial time to zero for simplicity, $t_0=0$.

\begin{figure}
\includegraphics[width=\columnwidth]{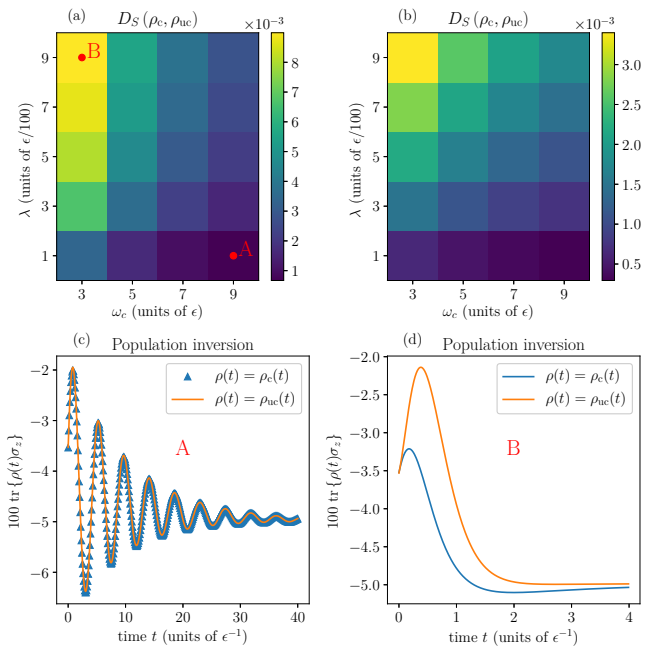}
\caption{\label{fig1} 
Demonstration of the systematic increase of the error introduced when approximating a correlated initial state by means of an uncorrelated initial state according to Eq.~(\ref{eq_init_prod_state}) in the strong-coupling and slow-environment regime. 
The top panels (a) and (b) depict the cumulative trace distance between solutions based on Eq.~(\ref{eq_init_state}) and Eq.~(\ref{eq_init_prod_state}), whereas the bottom panels (c) and (d) demonstrate the transient difference in observables directly.
(a), (c) and (d) Considering a correlated initial state obtained from a unitary transformation of the global thermal state according to Eq.~(\ref{eq_init_state}) with $R={e^{i\frac{\pi}{8}\sigma_x}}$. 
The panels (c) and (d) show the evolution of the population inversion of the two-level system for values of $\lambda$ and $\omega_c$ as indicated in panel (a). 
(b) Considering a correlated initial state obtained from a preparative measurement performed on the open system (see text). 
The initial inverse temperature for all plots is given by $\beta=0.1\epsilon^{-1}$, the considered tunneling amplitude is given by $\Delta=\epsilon$.
}
\end{figure}

It is often assumed that within the weak system-environment coupling regime, the effect of the preparative procedure Eq.(\ref{eq_init_state}) onto the environment can be safely neglected and the correlated initial states $\rho_\text{tot}(t_0)$ can consequently be well approximated by means of the usual initial product state assumption
\begin{align}
\rho_\text{tot}(0)\approx\rho(0) \frac{e^{-\beta H_E}}{Z_E}. \label{eq_init_prod_state}
\end{align}
Fig.~\ref{fig0} (a) provides a sketch of the considered setup. 
In order to verify that the error introduced with Eq.(\ref{eq_init_prod_state}) generally grows with the coupling between the open system and its environment as well as with characteristic environmental correlation times, we present numerical results obtained for a spin-boson model governed by the Hamiltonian
\begin{align}
H=\frac{\epsilon}{2}\sigma_z-\frac{\Delta}{2}\sigma_x +\sigma_x X+H_E.
\label{eq_spin-boson}
\end{align}
Here, $\epsilon$ denotes the energetic splitting of the two bare levels, $\Delta$ denotes the tunneling amplitude, 
and $X=\sum_k \gamma_k (a_k^\dagger+a_k)$ denotes the collective position of the bosonic environment described 
by $H_E=\sum_k \omega_k a_k^\dagger a_k$, with creation and annihilation operators $a_k^\dagger$ and $a_k$ for 
environmental modes $k$ with associated energies $\omega_k$. 
The coupling to the individual environmental modes is parametrized by $\gamma_k$. 
The usual Pauli matrices are denoted by $\sigma_\alpha$, the normalized eigenvectors 
of $\sigma_z$  are defined by 
$\sigma_z\ket{\uparrow}=+\ket{\uparrow}$ and $\sigma_z\ket{\downarrow}=-\ket{\downarrow}$. 
For the results presented in Figure~\ref{fig1}, characteristic bath correlation functions of the 
form $\text{tr}_E\left\lbrace e^{i H_Et}Xe^{-i H_Et} X e^{-\beta H_E}/Z_E\right\rbrace\approx\lambda\pi\left(\frac{1}{\beta}-i\frac{\omega_c}{2}\right) e^{-\omega_c t}$
with coupling strength $\lambda$ and reservoir cutoff $\omega_c$ have been employed. 
These correlation functions are suitable for the HEOM formalism and approximate the effect of a Drude-Lorentz bath, described by a spectral density $J(\omega)=\sum_k\gamma_k^2\delta(\omega-\omega_k)=\lambda\frac{\omega_c\omega}{\omega_c^2+\omega^2}$,  in the high-temperature regime \cite{ishizaki_09}. 
Thereby, the reorganization energy $\int_0^\infty d\omega J(\omega)/\omega=\lambda \pi/2$ quantifies the overall coupling strength between the two-level system and its environment and $\omega_c$ is inversely proportional to the characteristic correlation time of the latter.

The time evolution of the reduced open system state corresponding to the correlated initial condition Eq.(\ref{eq_init_state}), 
$\rho_c(t)=\text{tr}_E\left\lbrace e^{-iH t} \rho_\text{tot}(0) e^{iH t} \right\rbrace$, and the evolution of the reduced state 
corresponding to the uncorrelated initial condition Eq.(\ref{eq_init_prod_state}), 
$\rho_{uc}(t)=\text{tr}_E\left\lbrace e^{-iH t} \rho(0) \frac{e^{-\beta H_E}}{\text{tr}\left\lbrace e^{-\beta H_E} \right\rbrace} e^{iH t} \right\rbrace$, 
are compared in Fig.~\ref{fig1}. 
Their difference is quantified by means of the cumulative trace distance defined as 
$D_S(\rho_c,\rho_{uc})=\int_{0}^\infty dt~ D(\rho_{c}(t),\rho_{uc}(t))$, with $D(A,B)=\frac{1}{2}\text{tr}\left\lbrace\sqrt{(A-B)^\dagger(A-B)} \right\rbrace$ 
denoting the usual trace distance. 
From this definition one can already infer that the cumulative trace distance would diverge if differences between $\rho_c$ and $\rho_{uc}$ prevail at steady state.
The considered correlated initial states are given by means of a unitary rotation of the total system thermal state according to 
Eq.~(\ref{eq_init_state}) with $R={e^{i\frac{\pi}{8}\sigma_x}}$ (panel (a), (c) and (d)) or by means of an ideal measurement performed on the open system yielding an outcome associated to the open system state $\ket{\psi}=\frac{1}{\sqrt{2}}(\ket{\uparrow}+i\ket{\downarrow})$, such that the projector $P_o$ as described above is given by $P_o=\ket{\psi}\bra{\psi}$ (panel (b)). 
The upper plots (a) and (b) in Fig.~\ref{fig1} point out that the error introduced with Eq.~(\ref{eq_init_prod_state}) systematically increases when increasing the system-environmental coupling strength $\lambda$ or the characteristic correlation time of the environment $\omega_{c}^{-1}$. 
Panels (c) and (d) compare the effect of these initial correlations in terms of the time evolution of the population inversion for parameter sets A and B marked in panel (a).

%
%
%
%
\section{Initial correlations via the transfer tensor method}\label{SEC:init_corr_ttm}

In discrete time, the evolution of any reduced open system state can be expressed in terms of the transfer tensor formalism~\cite{cerrillo_tt} in the following generalized form
\begin{align}
\rho(t_n)=\sum\limits_{k=1}^n\mathcal{T}_k\rho(t_{n-k})+\mathcal{I}_n\left[\rho_\text{tot}(t_0)\right]\label{eq_dt_nz},
\end{align}
with transfer tensors $\mathcal{T}_k$ corresponding to the evolution of an initial product state and a correction 
$\mathcal{I}_n\left[\rho_\text{tot}(t_0)\right]$ that accounts for the effect of initial correlations between the system and its environment. 
Here, the transfer tensors are defined recursively by $\mathcal{T}_k=\mathcal{E}_k-\sum_{m=1}^{k-1}\mathcal{T}_{k-m}\mathcal{E}_m$ with $\mathcal{T}_1=\mathcal{E}_1$ 
and dynamical maps $\mathcal{E}_k$ that determine the evolution of an open system given an initial separable state of the form 
$\rho_\text{tot}(0)=\rho(0){e^{-\beta H_E}}/Z_E$, such that $\rho(t_k)=\mathcal{E}_k\rho(0)$. 
We emphasize that the tensors $\mathcal{T}_k$ carry key information about the underlying dynamics of the open system state as they manifest a discrete representation of the Nakajima-Zwanzig memory kernel within the reduced open system subspace.

\begin{figure}
\includegraphics[width=\columnwidth]{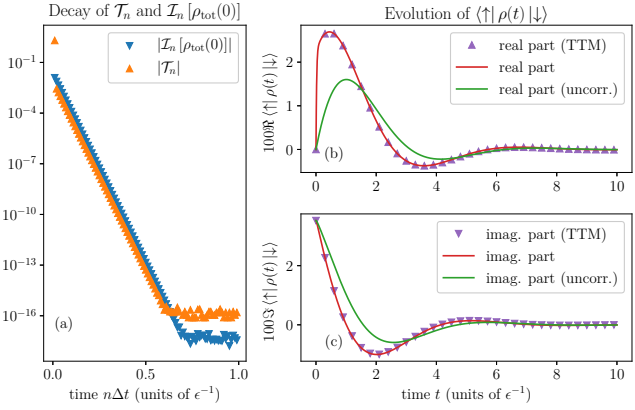}
\caption{\label{fig2} 
Demonstration of the application of the transfer tensor method for correlated initial states, 
considering the pure-dephasing spin-boson model Eq.~(\ref{Eq_dephasing}). 
The chosen parameters are $\beta=0.2\epsilon^{-1}$, $\lambda=0.5\epsilon$, $\omega_c=50\epsilon$
with a discrete time spacing of $\Delta t=0.01\epsilon^{-1}$. 
The considered correlated initial state reads as $\rho_\text{tot}(0)=R e^{-\beta H}/\text{tr}\left\lbrace e^{-\beta H} \right\rbrace R^\dagger$ 
with $R=e^{i \pi/ 8 \sigma_x}$. 
(a) Illustration of the decay of the exact transfer tensors $\mathcal{T}_n$ and the correction 
$\mathcal{I}_n\left[\rho_\text{tot}(0)\right]$ (using the Frobenius norm $| \bullet |$). 
(b) Evolution of the real part of the non-trivial reduced density matrix element $\bra{\uparrow}\rho(t)\ket{\downarrow}$. 
The corresponding result obtained when considering an uncorrelated initial state 
$\tra{E}{\rho_\text{tot}(0)} e^{-\beta H_E}/\text{tr}\left\lbrace e^{-\beta H_E} \right\rbrace$ 
is shown for comparison (green line). 
Triangles show the perfectly matching results obtained when using transfer tensors learned from the numerically exact HEOM 
for a subsequent propagation of the initial sample. The initial sample ranges from $t=0$ to $t=\epsilon^{-1}$ and covers the length 
of the memory kernel and the period of non-vanishing $\mathcal{I}_n\left[\rho_\text{tot}(0)\right]$. 
(c) Evolution of the imaginary part of the non-trivial reduced density matrix element, otherwise as (b).
}
\end{figure}

Here, we propose the use of transfer tensors for correlated initial states in the following manner:

\begin{itemize}

\item First, the transfer tensors corresponding to an initial product state featuring a thermal bath 
$\rho_\text{tot}(0)=\rho(0)e^{-\beta H_E}/Z_E$ are learned from a suitable simulation method of choice. 

\item In a second step, the reduced density matrix corresponding to the correlated initial state of interest is 
propagated up to some time $\tau>\omega_{c}^{-1}$ covering at least the length of the memory kernel, such that 
$\mathcal{K}(t>\tau-t_0)\approx 0$. 
This step can be conducted by means of any appropriate numerically exact simulation method, for instance by 
integrating the SLE~\cite{stockburger2002} or HEOM~\cite{tanimura_2014}.

\item Third, the correction term $\mathcal{I}_n\left[\rho_\text{tot}(0)\right]$ that quantifies the effect of initial 
correlations onto the dynamics can be determined. 
A sufficient decay of $\mathcal{I}_n\left[\rho_\text{tot}(0)\right]$ within the simulation time $\tau$ verifies 
the suitability of the final step: a subsequent efficient propagation by means of the transfer tensors only.
\end{itemize}

This approach is illustrated in Fig.~\ref{fig2} and benchmarked in combination with HEOM against exact results. 
For this, an exactly solvable variant of the spin boson model Eq.(\ref{eq_spin-boson}), given by 
\begin{align}
H=\frac{\epsilon}{2}\sigma_z+\sigma_z X+H_E \label{Eq_dephasing},
\end{align}
with a characteristic environmental correlation function 
$\text{tr}\left\lbrace e^{i H_E t}Xe^{-i H_E t} X e^{-\beta H_E}/Z_E\right\rbrace=\lambda\pi\left(\frac{1}{\beta}-i \frac{\omega_c}{2}\right) e^{-\omega_c t}$ is considered. 
The evolution of a correlated initial state 
$R e^{-\beta H}/Z R^\dagger$ with $R=e^{i\pi/ 8 \sigma_x}$ is shown in panels (b) and (c), 
by means of the real and imaginary part of the reduced density matrix element 
$\bra{\uparrow}\rho(t)\ket{\downarrow}$. 
The result obtained when considering an uncorrelated initial state 
$\tra{E}{\rho_\text{tot}(0)} e^{-\beta H_E}/Z_E$ is additionally shown for comparison. 
The large discrepancies between initially correlated and uncorrelated simulations are mostly due to the strong 
system-environment coupling and further reveal the necessity to account for initial correlations. 
Remarkable agreement is shown between the exact and the TTM-HEOM results. 
Panel (a) further verifies the suitability of the approach by showing the decay of the transfer tensors and 
the correction term $\left|\mathcal{I}_n\left[\rho_\text{tot}(t_0)\right]\right|$ within the range $t=0$ 
to $t=\epsilon^{-1}$. 
In view of the strong suppression over several orders of magnitude, the reduced open system state can be safely 
further propagated in time by means of transfer tensors as described above.

We emphasize that the verification of the suitability of the presented transfer tensor method for correlated initial 
states can be ensured by monitoring the required decays of the inhomogeneous contribution 
$\left|\mathcal{I}_n\left[\rho_\text{tot}(t_0)\right]\right|$ and the transfer tensors $\left|\mathcal{T}_n\right|$ -- 
even for models that do not allow for an exact solution. 
Sufficient decays have been verified for all of the presented results.

%
%
%
%
\section{Application: Emission spectra of multichromophoric systems}\label{SEC:emission_spectra}

As we will see, the calculation of emission spectra of multichromophoric systems provides an application that most importantly requires to account for correlated initial states. 
Here, we refer to  systems that have been extensively discussed in Refs.~\cite{cao2015forster_1, cao2015forster_2, cao2015forster_3}. 
The molecular systems of interest are modeled by means of $1\le i \le N$ internal excited states $\ket{e_i}$, 
that span a single-exciton manifold and that are coupled diagonally to the collective positions of their individual bosonic environments. 
Additionally, these systems feature an internal ground state, denoted by $\ket{g}$, that is uncoupled from the excited states and their 
thermal baths in the Hamiltonian description. 
However, transitions between the excited manifold and the internal ground state can be formally incorporated by interactions with a further
generic environment of the form $\mu \otimes B$, with system dipole operator
\begin{align}
\mu=\sum_{i=1}^N\left(\ket{e_i}\bra{g}+\text{H.c.}\right)
\end{align}
and some arbitrary environment operator $B$.
Omitting this further environment from the discussion, the corresponding Hamiltonian reads as
\begin{align}
H=&H_S+\sum_{i=1}^N\ket{e_i}\bra{e_i} X_i+H_E\label{eq_molecular_sys}
\end{align}
with a a total number of $N$ independent bosonic baths, 
$H_E=\sum_{i=1}^N\sum_{k_i}\omega_{k_i}a_{k_i}^\dagger a_{k_i}$, 
and their collective positions $X_i=\sum_{k_i}\gamma_{k_i}\left(a_{k_i}+a_{k_i}^\dagger\right)$. 
The open system Hamiltonian representing the single-exciton manifold reads as
\begin{align}
H_S=\sum_{i=1}^N\epsilon_i\ket{e_i}\bra{e_i}+\sum_{i<j=1}^N\left(v_{ij}\ket{e_i}\bra{e_j}+\text{h.c.}\right), \label{HSEM}
\end{align}
and allows for interactions $v_{ij}$ among the excited states. 
Emission $E(\omega)$ and absorption $A(\omega)$ spectra are generically linked to transitions between the 
internal excited states and the ground state. 
They are obtained by computing Fourier transforms of the respective dipole-dipole correlation functions~\cite{cao2015forster_1}, 
$E(\omega)=\int_{-\infty}^\infty dt~e^{-i\omega t}E(t)$ and $A(\omega)=\int_{-\infty}^\infty dt~e^{i\omega t}A(t)$ with
\begin{align}
A(t)&=\text{tr}\left\lbrace \mu(t) \mu \rho_\text{tot}^g(0) \right\rbrace,\\
E(t)&=\text{tr}\left\lbrace \mu(t) \mu \rho_\text{tot}^e(0) \right\rbrace .
\end{align}
Here, $\mu(t)=e^{i Ht}\mu e^{-i Ht}$ denotes the time-evolved dipole operator. 
The considered initial state for absorption processes $\rho_\text{tot}^g(0)$ is a product state composed of the internal ground state and thermal bath states. 
It can be obtained from the global thermal state by means of a projection onto the ground state, $\rho_\text{tot}^g(0)=R_g e^{-\beta H}/Z R_g^\dagger$. 
Here, $R_g$ denotes the normalized projector onto the ground state, $R_g=R_g^\dagger=P_g/\sqrt{\text{tr}\left\lbrace P_g e^{-\beta H}/Z \right\rbrace}$ with $P_g=\ket{g}\bra{g}$. 
For emission processes on the other hand, a correlated initial thermal state that is equilibrated within the excited subspace needs to be considered. 
It is defined as the projection of the global thermal state onto the excited subspace $\rho_\text{tot}^e(0)=R_e e^{-\beta H}/Z R_e^\dagger$, 
with $R_e=R_e^\dagger=P_e/\sqrt{\text{tr}\left\lbrace P_e e^{-\beta H}/Z \right\rbrace}$ and $P_e=\sum_{i=1}^{N}\ket{e_i}\bra{e_i}$. 
This state may be prepared by optical pumping~\cite{Kastler1967} or similar procedures.
We note that such transitions between the excited and the ground manifold are beyond Eq.~(\ref{eq_molecular_sys}) but would
e.g. correspond to the linear response effect of the system on additional environmental modes.

The absorption and emission spectra fulfill a Kubo-Martin-Schwinger like relation
\begin{align}
E(t-i\beta)=A^*(t)\frac{Z_E}{Z_e}\,,
\end{align}
with $Z_e=\text{tr}\left\lbrace P_e e^{-\beta H} \right\rbrace$, and which translates into the frequency domain as
\begin{align}
E(\omega)=A(\omega)e^{-\beta \omega}\frac{Z_E}{Z_e} \label{eq_temp_relation}.
\end{align}
This can be derived from the form of the dipole operators, $P_e\mu P_e=P_g\mu P_g=0$ and $P_e\mu P_g=P_g\mu P_e=1$, together with the fact that the total Hamiltonian does not couple the ground state with the excited manifold, $P_e H P_g=P_g H P_e=0$, and it fulfills $P_g H P_g=H_E P_g$ and $P_e H P_e=H P_e$. As a consequence, the absorption and emission
 dipole-dipole correlation functions can be rewritten as $E(t)=\sum_{i,j=1}^N\bra{e_i} \hat{E}(t)\ket{e_j}$ and
  $A(t)=\sum_{i,j=1}^N\bra{e_i} \hat{A}(t)\ket{e_j}$ with reduced absorption and emission operators
\begin{align}
\hat{A}(t)&=
\text{tr}_E\left\lbrace e^{-iHt}\frac{e^{-\beta H_E}}{Z_E}e^{i H_{E} t}\right\rbrace\label{eq_red_abs_op}
\intertext{and}
\hat{E}(t)&=
\text{tr}_E\left\lbrace e^{iHt}\rho_\text{tot}^e(0)e^{-i H_{E} t}\right\rbrace\label{eq_red_emi_op}.
\end{align}
These matrices fulfill the Kubo-Martin-Schwinger like relationship also elementwise.
Most importantly, relationship Eq.~(\ref{eq_temp_relation}) has proven useful for the simplification of the derivation of multicromophoric spectra \cite{Chenu2017} and can in our case be employed for temperature estimation from spectroscopic data and for the validation of results obtained from numerical simulations as shown in Figs.~\ref{fig4} and~\ref{fig5}.

\begin{figure}
\includegraphics[width=\columnwidth]{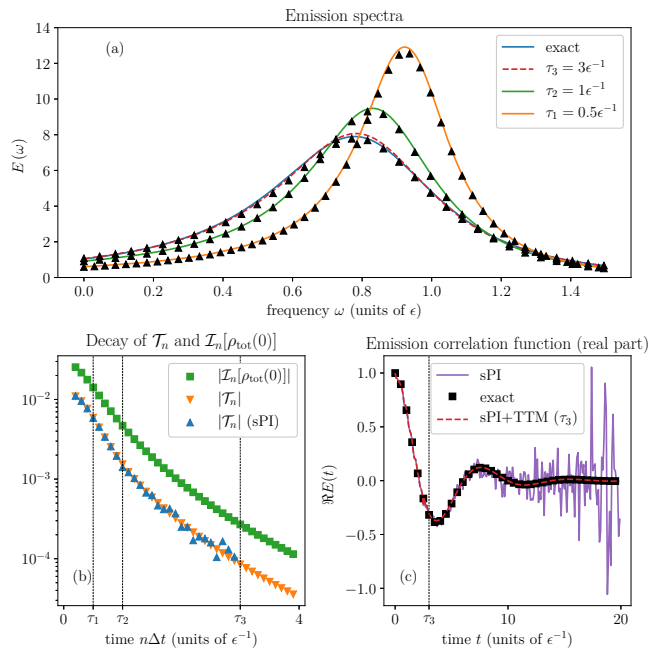}
\caption{\label{fig3} 
Application of the transfer tensor method for emission spectra of the exactly solvable model with $v=0$ and $\epsilon_1=\epsilon_2=\epsilon$. 
The environmental parameters are given by $\lambda=0.1\epsilon$, $\omega_c=2\epsilon$ and $\beta=\epsilon^{-1}$. 
Transfer tensors are obtained for a time discretization defined by $\Delta t=.1\epsilon^{-1}$. 
(a) Emission spectra are obtained from the analytic transfer tensor approach when considering different sample times 
$\tau$ (red dashed, green and yellow line) and from the TTM-sPI approach (black triangles). 
The analytic exact spectrum is shown for comparison (blue curve). 
(b) Decay of the inhomogeneous contribution $\mathcal{I}_n$ using the Frobenius norm $\left| \bullet\right|$. 
Additionally, the decay of the memory kernel corresponding to the initially uncorrelated object is illustrated by means of 
the norm of the analytic transfer tensors (orange triangles) and the corresponding tensors obtained from the stochastic simulation
(blue triangles) for comparison.
(c) Real part of the emission correlation function $E(t)$ as obtained from a sPI simulation (purple line) demonstrating the poor convergence for large simulation times. Black squares indicate the exact solution that can be perfectly recovered with the presented TTM-sPI approach (red dashed line), even for long times.
}
\end{figure}

In a series of previous papers~\cite{cao2015forster_1,cao2015forster_2,cao2015forster_3}, the simulation of the reduced absorption and emission operators by means of different perturbative methods, such as the full cumulant expansion and hybrid cumulant expansion, 
as well as numerically exact methods, such as the stochastic path integral formalism or HEOM and its stochastic extension~\cite{moix_hybrid_heom}, has been discussed in detail.
It turns out that, for general parameters, the stochastic path integral (sPI) approach~\cite{cao2015forster_3} manifests itself as a very powerful technique: 
it is not restricted to high temperatures or certain forms of the environmental spectral densities and it is suitable for the strong system-environment 
coupling regime. 
We point out that it constitutes a simpler method than SLE for density matrices~\cite{stockburger2002} due to the one-sided form of the evolution 
of the reduced absorption and emission operators, see Eq.~(\ref{eq_red_abs_op}) and Eq.~(\ref{eq_red_emi_op}). 
However, obtaining converged results for long-lasting system dynamics remains a cumbersome task. 
For the case of absorption spectra, this difficulty can be overcome by means of the usual TTM approach \cite{rosenbach_2016}. Here, we extend this result and address the emission operator with the presented TTM approach for correlated initial states.
We instantiate the proposed method with a molecular system that exhibits $N=2$ excited states and is described by the system Hamiltonian
\begin{align}
H_S=
\begin{pmatrix}
\epsilon_1&v&0\\
v^*&\epsilon_2&0\\
0&0&0
\end{pmatrix}
\label{Hmatrix}
\end{align} 
expressed in the basis $(\ket{e_1},\ket{e_2},\ket{g})$. 
Consequently, we consider two independent but identical baths that are coupled to the excited states. 
Each of them shall be characterized by an ohmic spectral density with an exponential cutoff 
$J(\omega)=\sum_{\omega_{k_i}}\gamma_{k_i}^2\delta\left(\omega-\omega_{k_i}\right)=\lambda\omega e^{-\omega/\omega_c}$ for $i=1,2$.

For a diagonal system Hamiltonian, $v=0$, the emission problem can be solved exactly and may be used for the systematic benchmark of 
our TTM-sPI approach for correlated initial conditions that is presented in Fig.~\ref{fig3}. Panel (a) shows the exact emission spectrum 
(blue line) and the rapid convergence of TTM towards the exact result with increasing sample times $\tau$ (red, green and yellow lines), 
where it is shown that a sample time of $\tau_3=3\epsilon^{-1}$ is sufficient to recover the exact result. 
The increase in the sample time gradually broadens the spectrum and its maxima is shifted to lower frequencies. 
This trend is representative of an effective increase of the system-environmental coupling as a larger portion of the memory kernel is represented by the transfer tensors. 
This view is supported by panel (b) which shows the decay of the transfer tensors (yellow triangles). 
It also shows the rapid decay of the effect of initial correlations (green squares) of approximately two orders of magnitude by the time $\tau_3$. 
Panel (c) illustrates the difficulty to access long time propagation of the emission correlation function by means of a sPI simulation (purple line), whereas transfer tensors learned from a short-time sPI simulation suffice to accurately propagate to long times (red dashed line).

\begin{figure}
\includegraphics[width=\columnwidth]{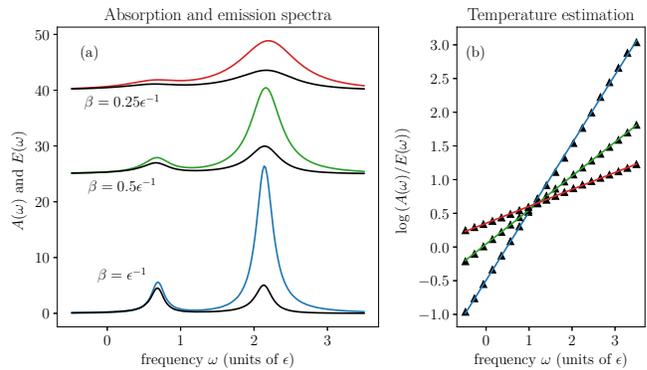}
\caption{\label{fig4}(a) Absorption (colored curves) and emission (black curves) spectra for three different temperatures are obtained by means of 
the TTM-sPI approach considering $\epsilon_1=2\epsilon$, $\epsilon_2=1\epsilon$ and $v=0.5\epsilon$. 
The same bath parameters for all three temperatures are given by $\lambda=0.05\epsilon$ and $\omega_c=2\epsilon$. 
For visibility, the spectra corresponding to $\beta=0.5\epsilon^{-1}$ ($\beta=0.25\epsilon^{-1}$) are vertically offset by $25$ ($40$). 
(b) Temperature estimation from the simulated emission and absorption spectra according to Eq.~(\ref{eq_temp_relation}). 
The slopes have been fitted to the simulated data by means of a least square method yielding  $\beta=1.016\epsilon^{-1}$, $\beta=0.504\epsilon^{-1}$ and $\beta=0.251\epsilon^{-1}$. Only every tenth data point (black triangles) is shown for the purpose of a clear presentation.}
\end{figure}

\begin{figure}
\includegraphics[width=\columnwidth]{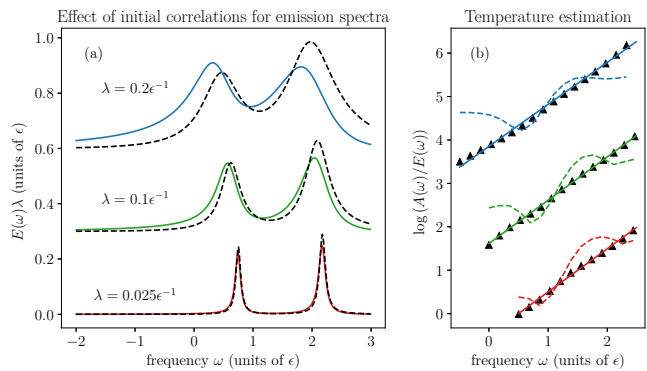}
\caption{\label{fig5} 
(a) Emission spectra for different system-environment couplings, simulated by means of the TTM-sPI approach when accounting for 
(solid curves) or neglecting (dashed curves) the effect of initial correlations. 
The system Hamiltonian is given by $\epsilon_1=2\epsilon$, $\epsilon_2=\epsilon$ and $v=0.5\epsilon$. 
The same bath parameters for all three values of $\lambda$ are given by $\beta=\epsilon^{-1}$ and $\omega_c=2\epsilon$. 
The spectra corresponding to $\lambda=0.1\epsilon$ ($\lambda=0.2\epsilon$) are vertically offset by $0.3\epsilon$ ($0.6\epsilon$). 
(b) Temperature estimation from the simulated emission (when accounting for initial correlations) and absorption spectra 
according to Eq.~(\ref{eq_temp_relation}). 
The slopes have been fitted to the simulated data by means of a least square method yielding  
$\beta=0.994\epsilon^{-1}$ ($\lambda=0.025\epsilon$), $\beta=0.997\epsilon^{-1}$ ($\lambda=0.1\epsilon$) and $\beta=0.968\epsilon^{-1}$ ($\lambda=0.2\epsilon$). 
Only every fourteenth data point is shown for the purpose of a clear presentation. 
The data points corresponding to $\lambda=0.1\epsilon$ ($\lambda=0.2\epsilon$) are vertically offset by $2$ ($4$). 
Dashed lines are obtained when considering the emission spectra that have been obtained by neglecting the effect of initial correlations.}
\end{figure}

The symmetry relation between absorption and emission spectra Eq.(\ref{eq_temp_relation}) allows for thermometry based on spectroscopic data and also 
for the validation of the results stemming from numerical simulations such as our joint TTM-sPI approach. 
By means of this relation, the environmental temperature as well as the ratio of the partition functions 
$Z/Z_E=\text{tr}\left\lbrace e^{-\beta H}\right\rbrace/\text{tr}\left\lbrace e^{-\beta H_E}\right\rbrace$ 
can be immediately recovered from the corresponding spectra as illustrated in Fig.~\ref{fig4}.
Therein, we consider the Hamiltonian
\begin{align}
H_S=
\epsilon
\begin{pmatrix}
2&0.5&0\\
0.5&1&0\\
0&0&0
\end{pmatrix}.
\end{align}
In panel (a) emission and absorption spectra are shown for three different temperatures 
$\beta^{-1}=\epsilon$ (low), $\beta^{-1}=2\epsilon$ (intermediate) and $\beta^{-1}=4\epsilon$ (high). 
All of the presented emission (black lines) and absorption (colored lines) spectra show peaks at frequencies that can 
be roughly identified with the two eigenvalues of the system Hamiltonian given by $2.207\epsilon$ and $0.793\epsilon$, 
which can be best recognized in the low temperature example. 
The intermediate and high temperature cases exhibit an enhanced occupation of energetically higher environmental states, 
which supports transitions between ground and the excited states and vice versa on a wider energetic range. 
This results in a general broadening of the spectra. 
Additionally, one finds that for the low temperature example, the two emission peaks show roughly the same height, 
whereas for increasing temperatures one observes an increasing imbalance in favor of the peak associated to the higher energy. 
Panel (b) shows that successful estimation of the temperature from the simulated data is possible when initial correlations are taken into account.

The presented TTM approach for correlated initial conditions allows for detailed studies of the importance of initial correlations 
for emission spectra by means of the possibility of accounting fully or partially, or even neglecting, the corresponding effects. 
In practice, this can be achieved by truncating the length of the initial sample of the exact (initially correlated) evolution of 
$\hat{E}(t)$ that is used. 
This is shown in Fig.~\ref{fig5} (a), where approximate emission spectra are presented which are obtained from a simulation that 
neglects the effect of initial correlations by means of a propagation of a sample of $\hat{E}(t)$ that only has length one. 
For weak system-environment couplings, one expects the effect of initial correlations to be negligible and this is indeed 
the case for $\lambda=0.025\epsilon$, for which the approximate emission spectrum (black dashed line) provides a qualitatively 
good estimate for the exact emission spectrum (red line). 
However, with increasing system-environmental coupling strength, the effect of initial correlations becomes more relevant for 
the reduced emission operator and the approximate spectra fail to estimate the exact result. 
More importantly, we stress that even for the weak-coupling case ($\lambda=0.025\epsilon$) the temperature cannot be faithfully 
recovered from the approximate spectrum as shown in panel (b). 
Temperature estimation from the simulated spectra crucially requires to account for initial correlations of the reduced 
emission operator $\hat{E}(t)$.

%
%
%
%
\section{Application: Spectral signatures of electromagnetically induced transparency}\label{SEC:darkstates}

We finally employ the joint TTM-sPI approach for the investigation of spectral signatures of electromagnetically induced transparency (EIT). EIT is the phenomenon of vanishing absorption or emission due to the coupling of the dissipative manifold to a discrete state, and has found extensive application in the enhancement of laser cooling schemes for trapped ions~\cite{Morigi2000}.
It appears in a so-called lambda configuration, where an excited state $\ket{e}$ is coupled in Raman resonance with two ground states $\ket{g_1}$ and $\ket{g_2}$ forming the following system Hamiltonian
\begin{align}
H_S=\epsilon\ket{e}\bra{e}+\sum_{i=1}^2\left(v_{i}\ket{e}\bra{g_i}+\text{h.c.}\right),
\label{HSDS}
\end{align}
For the case $v_1=v_2$, the basis of the dark state $\ket{-}=\frac{1}{\sqrt 2}\left(\ket{g_1}-\ket{g_2}\right)$ and bright state $\ket{+}=\frac{1}{\sqrt 2}\left(\ket{g_1}+\ket{g_2}\right)$ allows us to represent Hamiltonian Eq.(\ref{HSDS}) in the form Eq.(\ref{Hmatrix}). We will investigate the specific case
\begin{align}
H_S=\epsilon
\begin{pmatrix}
6&1 &0\\
1&2 &0\\
0&0 &0\\
\end{pmatrix}
\label{HSDS}
\end{align}
where the matrix is represented in the $(\ket{e},\ket{+},\ket{-})$ basis. 
\begin{figure}
\includegraphics[width=\columnwidth]{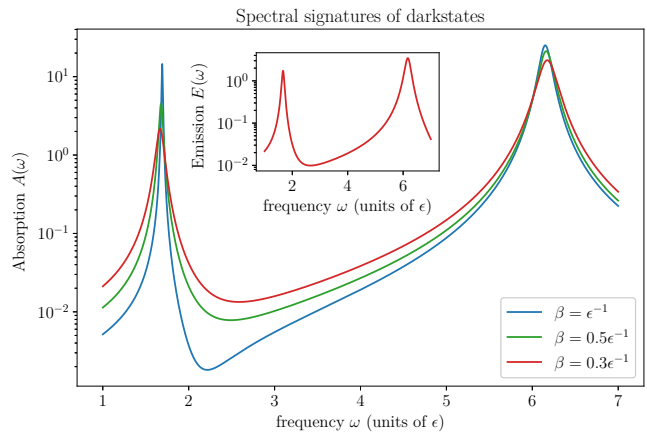}
\caption{\label{fig6} Demonstration of spectral signatures of darkstates considering $\omega_c=10\epsilon$ and $\lambda=0.01 \epsilon$ and Hamiltonian Eq.~(\ref{HDS}). The location of the peaks approach the eigenenergies of Hamiltonian Eq.~(\ref{HSDS}), whereas the dip approaches for low temperatures the energetic value of the bare bright state $\ket{+}$ at $\omega\simeq 2\epsilon$.}
\end{figure}
This three-level system is coupled to the collective position $X=\sum_{k}\gamma_{k}\left(a_{k}+a_{k}^\dagger\right)$ of a 
bosonic environment $H_E=\sum_{k}\omega_{k}a_{k}^\dagger a_{k}$ in dipolar form, such that the total Hamiltonian of the considered setup reads as
\begin{align}
H=H_S+
\begin{pmatrix}
0&1&0\\1&0&0\\0&0&0
\end{pmatrix}
X
+H_E.
\label{HDS}
\end{align}
The coupling to the individual environmental modes shall be parametrized by an Ohmic spectral density with an 
exponential cutoff $J(\omega)=\sum_{\omega_k}\gamma_{k}^2\delta\left(\omega-\omega_{k}\right)=\lambda\omega e^{-\omega/\omega_c}$. 
We focus on the corresponding emission and absorption spectra for the dipole operator definition $\mu'=\left(\ket{e}\bra{+}+\text{H.c.}\right)$
that show signatures of electromagnetically induced transparency. 
Absorption spectra are shown in Fig.~\ref{fig6} when considering three different temperatures of the environment. 
The characteristic dip indicating the presence of a dark state at $\omega\simeq 2\epsilon$ becomes most obvious for low temperatures of the environment. However, even for the highest considered temperature $\beta=0.3\epsilon^{-1}$ the characteristic suppression can be observed within the corresponding emission spectrum.

We note that the presented TTM-sPI approach is especially well suited for studies of the parameter-regime considered in Fig \ref{fig6}. This is because of the long-lasting system dynamics due to the weak system-environmental coupling and the low temperatures. The considered parameter regime could otherwise not be treated with the usual stochastic path integral method.

%
%
%
%
\section{Summary}
Successful TTM-sPI and TTM-HEOM approaches for the simulation of open quantum systems subject to correlated initial conditions have been presented in this paper. 
The TTM approach for correlated initial conditions is especially useful for settings that feature low-temperatures or intermediate system-environment couplings, 
such that the reduced open system state of interest evolves on a time scale that is large compared to the characteristic correlation time of the environment.

We note that originally, TTM has been introduced for open systems that are subject to uncorrelated initial conditions. For the class of correlated initial conditions considered in this paper, it is still possible to apply TTM without any significant overhead.
Recently, it has been pointed out that the formalism can in principle be extended in order to include correlated initial states by expressing them as a linear combination of a maximum of $d^2$ system-environment product terms~\cite{pollock2017}, where $d$ denotes the open system Hilbert space dimension. 
For each of the terms, the original TTM can be applied individually, leading to a set of $d^2$ transfer tensors that propagate the individual terms. 
Nevertheless, a practical implementation of the concept remains pending and the required overhead involving the derivation of $d^2$ transfer tensors could render it a challenging task.

Considering absorption and emission spectra of multichromophoric systems, we have shown that the symmetry relation Eq.~(\ref{eq_temp_relation}) allows for 
successful temperature estimation from simulated spectroscopic data. 
In this context it turned out to be crucial to account for initial correlations of the reduced emission operator -- even in the weak system-environment coupling regime.

Finally, the joint TTM-sPI approach enables the exploration of spectral signatures of electromagnetically induced transparency within obtained absorption and emission spectra in hardly accessible regimes.

\acknowledgements
The authors would like to acknowledge financial support from the DFG projects BR 1528/9-1, SCHA 1646/3-1, SFB 910, GRK 1558. M.B. acknowledges financial support from the Studienstiftung des Deutschen Volkes.

%
%
%
%
\bibliography{init_corr_bib}

\end{document}